\documentclass[aps,prb,twocolumn,showpacs]{revtex4}

\usepackage{graphicx}

\begin{document}

\bibliographystyle{apsrev}

\title{Thermodynamic driving force of formation of coherent three-dimensional
islands in Stranski-Krastanov growth}

\author{Jos\'e Emilio Prieto}
\altaffiliation{Present address: Institut f\"ur Experimentalphysik, Freie
Universit\"at Berlin, Arnimallee 14, 14195 Berlin, Germany}
\affiliation{Departamento de F\'\i{}sica de la Materia Condensada and
Instituto Universitario de Ciencia de Materiales ``Nicol\'as Cabrera",
Universidad Aut\'onoma de Madrid, 28049 Madrid, Spain}

\author{Ivan Markov}
\email{imarkov@ipchp.ipc.bas.bg}
\affiliation{Institute of Physical Chemistry, Bulgarian Academy of Sciences,
1113 Sofia, Bulgaria}
\date{\today}

\begin{abstract}
The formation of coherent three-dimensional islands in highly mismatched 
epitaxy is discussed in terms of the traditional concept of wetting. It is
shown that the wetting layer and the 3D islands represent different phases
which cannot be in equilibrium with each other. The transfer of matter
from the stable wetting layer to the 3D islands is thermodynamically
unfavored. The experimentally observed critical misfit for coherent 3D
islanding to occur and the coexistence of pyramids with discrete heights of
two, three, four... monolayers can be explained assuming that the thermodynamic
driving force of formation of coherent 3D islands on the surface of the
wetting layer of the same material is the reduced average adhesion of the
islands to that layer and that the islands height is a discrete variable.
\end{abstract}


\maketitle

The growth of thin epitaxial films usually takes place far from 
equ\-i\-lib\-ri\-um. Nevertheless,
ther\-mo\-dy\-na\-mic con\-si\-der\-a\-tions are a necessary step for
un\-der\-stand\-ing of the pro\-cess. Of particular interest is the
ther\-mo\-dy\-na\-mic driving force (TDF) which is responsible for one or 
another
mechanism of growth. While this ques\-tion is well understood in terms of
wet\-ting of the substrate by the overgrowth in the cases of island or
Vol\-mer-Weber (VW) growth and layer-by-layer or Frank-van der Merwe (FM)
growth,\cite{Bauer,Kern,Gilmer} the Stranski-Krastanov (SK) growth (3D
islands on top of a thin  wetting layer) is far from being clarified. The
reason is that the SK growth is in fact a growth of material A on the same
material A, which thermodynamically requires the formation and growth of 2D
rather than 3D islands. This is particularly true in the case of the
{\em coherent}
Stranski-Krastanov growth,\cite{Eagle} where dislocation-free 3D islands are
strained to the same degree as the wetting layer.\cite{Eagle,Mo,Moison} This
is the reason why it is widely accepted that the energy of the interfacial 
boundary
between the 3D islands and the wetting layer is equal to zero.\cite{J1}
Although this energy is expected to be small compared with that of the
free crystal faces,\cite{Jacques} it should not be neglected since this is
equivalent to the assumption that the islands wet completely the wetting
layer. The latter rules out 3D islanding from a thermodynamic
point of view.\cite{Sir}

The need for a thermodynamic anal\-y\-sis arises also from
the experimental observations of a critical misfit for coherent 3D islanding
to occur,\cite{Xie,Leo,Walther,Pinc} and the si\-mul\-ta\-neous presence of
islands of different thickness which vary by one monolayer.\cite{Rud,Col} The
existence of a critical misfit, as well as of stable two, three or four
monolayers thick islands, do not follow from the tradeoff\cite{J1,Jacques}
\begin{eqnarray}\label{energy}
\Delta E \approx C^{\prime }\gamma V^{2/3} -
C^{\prime\prime}\varepsilon _{0}^{2}V
\end{eqnarray}
be\-tween the cost of the additional surface energy and the gain of energy
due to the elastic relaxation of the 3D islands relative to the wetting layer
($V$, $\gamma $ and $\varepsilon _{0}$ are the islands volume, the specific
surface energy and the lattice misfit, respectively, and $C^{\prime }$ and
$C^{\prime\prime}$ are constants).

It was recently suggested that the TDF for coherent 3D is\-land\-ing is the
incomplete wetting of the substrate by the islands,\cite{Kor} rather than the
elastic relaxation of the ma\-te\-ri\-al in the islands. The incomplete
wetting is due to the displacements of the atoms near the island
edges from the bottoms of the corresponding potential troughs provided by the
wetting layer. This results in a series of critical volumes at which the
monolayer high islands become unstable against the bilayer islands, the
bilayer islands against the trilayer islands, etc. The misfit dependence of
the first critical size $N_{12}$ for the mono-bilayer transformation displays
a critical behavior in the sense that coherent 3D islands can be formed at a
misfit higher than some critical value. Below this value the film should grow
in a layer-like mode until misfit dislocations are introduced to relieve the
strain. However, the ap\-pro\-xi\-ma\-tion used by the authors, which is based
on the 1D model of Fren\-kel and Kontorova,\cite{Ratsch,Frenkel} was unable
to describe correctly the individual behavior of atoms inside each layer,
since it assumes a potential with a period given by the average of the
separations of atoms (considered frozen) in the layer underneath.
Although this model gives qua\-li\-ta\-ti\-ve\-ly reasonable results 
concerning the energy of the islands, it is inadequate to calculate, 
in particular, the average adhesion energy of the islands to the 
wet\-ting layer.

In the present report we recollect some simple ther\-mo\-dy\-na\-mic as\-pects
of the epitaxial mor\-pho\-lo\-gy based on the tra\-di\-ti\-o\-nal concept of
wet\-ting and consider the coherent SK growth from this point of view. The
same concepts were in fact advanced by Stranski in his model, admittedly very
peculiar,
of a monovalent ionic crystal K$^+$A$^-$ on the surface of an isomorphous
bivalent crystal K$^{2+}$A$^{2-}$.\cite{Stran1} We then support our 
thermodynamic con\-si\-der\-a\-tions by nu\-me\-ri\-cal cal\-cu\-la\-tions
making use of a simple mi\-ni\-mi\-za\-tion procedure on the same atomistic
model in $1+1$ di\-men\-sions (length + height) as in Ref.(\onlinecite{Kor}).
The 3D islands are represented by linear chains of atoms stacked one upon the
other,\cite{Ratsch} the islands height being thus considered as a discrete
variable which increases by unity from one. The latter is of crucial
importance as the aspect ratio of the 3D
islands is usually of the order of 0.1 and the height is of the order of 10 
monolayers.\cite{Mo,Moison} The atoms interact through an anharmonic Morse 
potential
\begin{equation}\label{potent}
V(x) = V_{o}[e^{-12(x-b)} - 2e^{-6(x-b)}].
\end{equation}
The total interaction energy as well as its derivatives with respect to
the atomic coordinates, i.e. the forces, are calculated. Relaxation is then 
performed iteratively by allowing the atoms to displace in the direction 
of the forces until these fall below some negligible cutoff value.
We consider interactions in the first coordination sphere in order to
mimic the directional bonds that are characteristic for
semiconductors.\cite{Tersoff} Inclusion of further coordination spheres
alters only minimally the numerical results. The substrate (the wetting 
layer) is assumed to be rigid.

\begin{figure}[h]
\center
\includegraphics*[height=4.5cm]{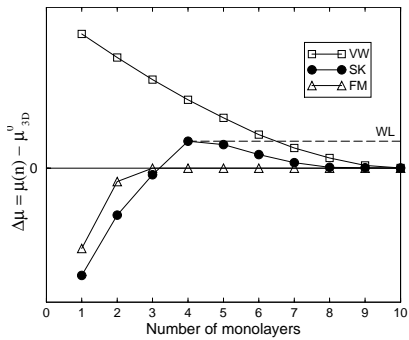}
\caption{\label{chempot} \small
Schematic dependence of the film chemical pot\-ent\-ial on the film thickness
in number of monolayers for the three modes of growth:
VW -- Volmer-Weber, SK -- Stranski-Krastanov, and FM -- Frank-van der Merwe.
The dashed line gives the chemical potential of the unstable wetting layer.}
\end{figure}

A mother phase (a vapor) and a new phase (e.g. a strained planar film or
unstrained 3D crystals) are in equilibrium with each other when their chemical
potentials are equal. A transition from one phase (mother or new) to another takes
place when the chemical potential of one of the phases becomes smaller than
that of the other. The TDF for this transition is the dif\-fer\-ence of the
chemical potentials of both phases at the given pressure and temperature. The
TDF which determines the occurrence of one or another mechanism of epitaxial
growth (growth from vapor of a strained 2D layer or 3D islands) is the
difference
$\Delta \mu =\mu (n)-\mu _{\rm 3D}^{0}$ of the chemical potential, $\mu (n)$, of
the overlayer which depends on the film thickness measured in number $n$ of
monolayers counted from the interface, and the chemical potential,
$\mu _{\rm 3D}^{0}$, of the bulk 3D crystal of the same
material.\cite{Kern,Gilmer} The thickness dependence of $\mu (n)$ originates
from the thickness distribution of the misfit strain and, on the other hand, 
from the interaction between the deposit and the substrate, which rapidly 
decreases with the distance from the interface 
($E_{\rm AB} \rightarrow E_{\rm AA}$).\cite{Kern,Mark1}

If we deposit a crystal A on the surface of a crystal B, $\Delta \mu $ can be
written in terms of the in\-ter\-a\-to\-mic energies per atom, $E_{\rm AA}$ and
$E_{\rm AB}$, required to disjoin a half-crystal A from a like half-crystal
A and from an unlike half-crystal B, respectively:\cite{Kaisch}
\begin{equation}\label{a-b}
\mu (n)=\mu _{\rm 3D}^{0}+[E_{\rm AA}-E_{\rm AB}(n)] =
\mu _{\rm 3D}^{0}+E_{\rm AA}\it \Phi .
\end{equation}

The adhesion energy $E_{\rm AB}$ includes in itself the thick\-ness
distribution of the strain energy and the at\-te\-nu\-a\-tion of the bonding
with the substrate. ${\it \Phi } = 1-E_{\rm AB}/E_{\rm AA}$ is the
{\it adhesion parameter} which accounts for the wetting of the substrate by
the overgrowth. Eq. (\ref{a-b}) is equivalent to the familiar 3-$\sigma $
criterion of Bauer.\cite{Bauer,Mark1}

As follows from (\ref{a-b}) the parameter ${\it \Phi } =
\Delta \mu /E_{\rm AA}$ is equal to the TDF for occurrence of one or another
mode of growth relative to the cohesive energy $E_{\rm AA}$. In the two
limiting cases of VW ($0 < {\it \Phi } < 1$) and FM growth (${\it \Phi } \le
0$, $\varepsilon _{0} \approx 0$), $\Delta \mu $ tends asymptotically with
increasing film thickness to zero from above and from below, respectively,
but changes its sign in the case of SK growth (${\it \Phi } < 0$,
$\varepsilon _{0} \ne 0$), as shown in
Fig.\ \ref{chempot}.\cite{Gilmer,Mark1} Consider now Fig.\ \ref{chempot}
in terms of equilibrium vapor pressures instead of chemical potentials.
Although the con\-nec\-tion is straightforward,
($\mu  \propto \ln P$), such a con\-si\-der\-a\-tion gives a deeper insight
into the problem.\cite{Stran1} Thus, as long as
$\mu (n) < \mu _{\rm 3D}^{0}$ a thin planar film can be
deposited at a vapor pressure $P$ that is smaller than the equilibrium vapor
pressure, $P_{0}$, of the bulk crystal, but is larger than the equilibrium
vapor pressure $P_{1}$ of the first monolayer, i.e. $P_1 < P < P_0$. In other
words, a planar film can be deposited at {\it undersaturation} $\Delta \mu  =
kT\ln (P/P_0)$ with respect to the bulk crystal. The formation
of 3D islands ($\mu (n) > \mu _{\rm 3D}^{0}$) requires $P > P_0$, or
a {\it su\-per\-sa\-tu\-ra\-tion} with respect to the bulk crystal.

Applying the above considerations to the SK growth leads unavoidably to the
conclusion that the 3D islands and the wetting layer represent
nec\-es\-sar\-ily different phases and thus have different chemical
potentials. The reason is that the two phases are in equ\-i\-lib\-ri\-um with
the mother phase (the vapor) under different conditions which never overlap.
The wetting layer can be in equ\-i\-lib\-ri\-um only with an
un\-der\-sa\-tu\-rated vapor phase ($P < P_0$), while small 3D islands can
be in equilibrium only with a su\-per\-sa\-tu\-ra\-ted vapor phase ($P >
P_0$). The dividing line is thus $\Delta \mu  = kT\ln (P/P_0) = 0$ at which
the wetting layer cannot grow thicker and the 3D islands cannot nucleate and
grow. Hence the wetting layer and the 3D islands can never be in equilibrium
with each other.

It follows from the above that the derivative, $d\Delta E/dV$, of the energy 
of the 3D islands relative to that of the wetting layer, gives the difference
of the chemical potentials of the wetting layer (the dashed line in Fig.\
\ref{chempot}) and the chemical potential of the 3D islands. In other words,
it represents the difference of the supersaturations of the vapor phase with
respect to the wetting layer and the 3D islands. As the thick\-ness and the
energy of the wetting layer depend on the misfit it would be more suitable if
one chooses as a ref\-er\-ence the bulk crystal rather than the wetting
layer.\cite{Muller} Transfer of material from the stable wetting layer
($\mu _{\rm WL} < \mu _{\rm 3D}^{0}$) to the 3D islands is connected with
increase of the free energy of the system, and therefore, is thermodynamically
unfavored. A planar film thicker than the stable wetting layer is unstable and
the excess of the material can be transferred to the 3D islands if the
necessary thermal activation exists.

\begin{figure}[h]
\center
\includegraphics*[height=4.5cm]{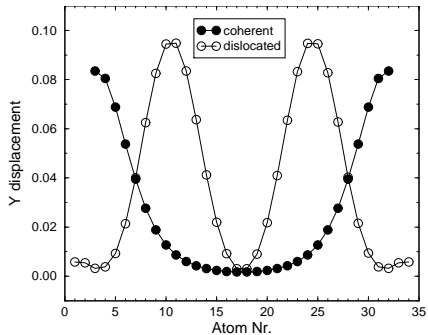}
\caption{\label{displ} \small
Vertical displacements of the atoms of the base chain of a coherent (a)
and a dislocated (b), 3 monolayers-thick island, given in units of the
lattice parameter of the wetting layer and measured from the bottoms
of the potential troughs provided by the homogeneously
strained wetting layer.
The misfit $\varepsilon_{0}$ amounts to 7\% and the islands contain
30 and 34 atoms in their base chains, respectively.}
\end{figure}

\begin{figure}[h]
\center
\includegraphics*[height=4.5cm]{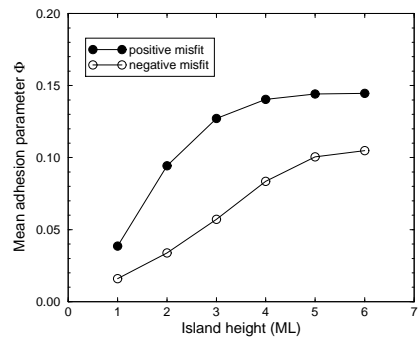}
\caption{\label{phi} \small
Mean adhesion parameter $\it \Phi $ as defined by Eq. (1) as a function of
the islands' height in number of monolayers for positive and negative
values of the misfit of absolute value of 7\%. Coherent islands
of 14 atoms in the base chain were considered in the calculations.}
\end{figure}

We focus our attention on the adhesion between the 3D islands and the wetting
layer. Fig.\ \ref{displ} is an illustration of the difference (and
resemblance) between the classical and the coherent SK mode. 
In the calculations, the sizes of the base chain of the island 
(30 and 34 atoms, respectively) have been chosen just below and above 
the critical size for introduction of misfit dislocations at the 
given misfit of 7\%.
As seen, in both cases the 3D islands loose contact with the wetting layer. 
The vertical displacements are largest at the chain's ends in the coherent 
SK mode and around the dislocation cores in the classical case, but the 
physics is essentially the same. The mean adhesion parameter $\it \Phi $
increases with the islands' height and saturates beyond several monolayers
(Fig.\ \ref{phi}). In our model, $\it \Phi $ is calculated as the adhesion
energy between island and wetting layer at the given misfit (7\%) minus the
corresponding value for zero misfit. It can be seen that the compressed 
overlayers exhibit a greater tendency to coherent SK growth than expanded 
ones as expected, due to the anharmonicity of the potential (\ref{potent}).

\begin{figure}[h]
\center
\includegraphics*[height=4.5cm]{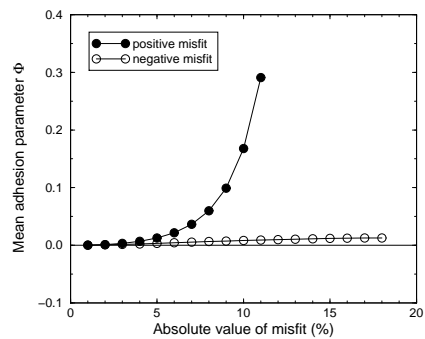}
\caption{\label{phif} \small
Mean adhesion parameter of one monolayer high, coherent islands as a function
of the lattice misfit. The islands contain 20 atoms. Data for both positive
and negative misfits are shown in one quadrant for easier comparison.}
\end{figure}

\begin{figure}[h]
\center
\includegraphics*[height=4.5cm]{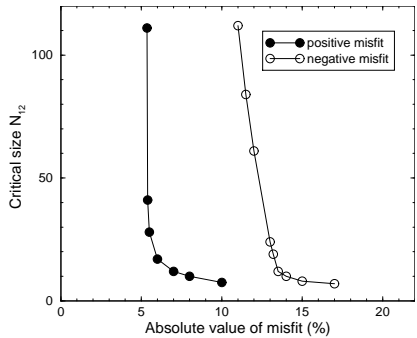}
\caption{\label{N12} \small
Misfit dependence of the critical size $N_{12}$ (in number of atoms) for
positive and negative values of the lattice misfit. The curves are shown
in one quadrant for easier comparison.}
\end{figure}

The same tendency is seen more clearly in Fig.\ \ref{phif} which shows in fact
the dependence of the TDF for formation of coherent 3D islands, $\Delta \mu$,
on the lattice misfit. The latter remains close to zero for expanded overlayers
but increases steeply beyond approximately 5\% in compressed overlayers. This
be\-ha\-vi\-or agrees well with the misfit dependence of the critical size,
$N_{12}$, for the mono-bilayer transformation to occur, as shown in Fig.\
\ref{N12}, where a steep rise of $N_{12}$ with decreasing absolute value of
the misfit is observed only in compressed over\-lay\-ers. Note that it is less
sharp for expanded ones opposite to earlier finding.\cite{Kor}

We discuss now the discrete character of the height of the 3D islands. The
experimentally observed volume of the quantum dots varies roughly from 20000
to 50000 atoms.\cite{Moison,Leo} Typical values of the aspect ratio of the
islands height and half-base are of the order of 0.1.\cite{Moison,Voigt} Thus 
a pyramid with a base edge of 100 atoms and aspect ratio 0.1, and containing
22000 atoms, is only 5 monolayers high. The addition of 14400 atoms
(a new base plane of $120\times120$ atoms) requires only one more atomic
plane. Calculations of the energy of islands having a shape of a frustum of a
pyramid are usually performed assuming implicitly that the lengths of the
lower, $R$, and the upper, $R^{\prime}$, bases, and in particular the height
$h$, are continuous variables. Eq. (\ref{energy}) is obtained by using the
Tersoff approximation which neglects the gradient of strain in a direction
normal to the surface plane together with $h \ll R$,\cite{J1} and assuming
$h \gg c$ where $c$ is the atomic distance.\cite{Jacques} This would be
correct if the crystals contain at least several million of atoms.

We conclude that the wetting layer and the 3D islands represent different
phases which cannot be in equilibrium with each other, and the SK morphology
is a result of the replacement of one first order phase transition (vapor --
wetting layer) by another first order transition (vapor -- 3D islands). The
transfer of mat\-ter from the stable wetting layer to the 3D islands is
thermodynamically unfavored. The experimental observations men\-ti\-oned above
can be explained on the base of two as\-sump\-tions: the thermodynamic driving
force for the coherent 3D islanding is the incomplete wetting and the height
of the 3D islands is a discrete variable varying by one monolayer. This leads
to the results that (i) monolayer high islands with a critical size appear as
necessary precursors for 3D islands, (ii) the 2D-3D transition takes place
through a series of intermediate states with discretely increasing thickness
that are stable in separate intervals of volume (see Refs.
(\onlinecite{Rud,Col})) (iii) there exists a critical misfit below which
coherent 3D islands are thermodynamically unfavored (see Refs.
(\onlinecite{Xie,Leo,Walther,Pinc})) and the misfit is accommodated by misfit
dislocations at a later stage of the growth. Compressed overlayers show
a greater tendency to 3D clustering than expanded ones, in agreement
with experimental results.\cite{Xie} Result (i) explains readily why
the volume distribution of InAs/GaAs self-assembled quantum dots agrees well
with the scaling functions for two-dimensional submonolayer homoepitaxy
model.\cite{Ebiko}

The authors are indebted to the Instituto Universitario de Ciencia de
Materiales ``Nicol\'as Cabrera" for granting research visits which enabled
scientific collaboration. This work was supported by the Spanish CICyT
through project Nr. MAT98-0965-C04-02.

\bibliography{apsrev}

\end{document}